\documentclass[aps,onecolumn,superscriptaddress,groupedaddress,nofootinbib]{revtex4}
\usepackage{graphicx}
\usepackage{dcolumn}
\usepackage{bm}
\usepackage{mathrsfs}
\usepackage{hyperref}
\usepackage{amsmath}
\usepackage{amssymb}
\usepackage{bm}
\usepackage{color}
\usepackage{float}
\usepackage{dcolumn}
\usepackage{multirow}
\usepackage{changepage}
\usepackage{enumerate}
\usepackage{setspace}
\hypersetup{pdftex,colorlinks=true,linkcolor=blue,citecolor=red,menucolor=black,urlcolor=blue,filecolor=blue}

\newcommand{\fm}{\textrm{ fm}}
\newcommand{\mev}{\textrm{ MeV}}

\begin{document}

\title{Evolution of genuine  states to molecular ones: The $T_{cc}(3875)$ case }

\author{L. R. Dai}
\email[]{dailianrong@zjhu.edu.cn}
\affiliation{School of Science, Huzhou University, Huzhou 313000, Zhejiang, China}
\affiliation{Departamento de F\'{\i}sica Te\'orica and IFIC, Centro Mixto Universidad de Valencia-CSIC Institutos de Investigaci\'on de Paterna, Aptdo.22085, 46071 Valencia, Spain}

\author{J. Song}
\email[]{Song-Jing@buaa.edu.cn}
\affiliation{School of Physics, Beihang University, Beijing, 102206, China}
\affiliation{Departamento de F\'{\i}sica Te\'orica and IFIC, Centro Mixto Universidad de Valencia-CSIC Institutos de Investigaci\'on de Paterna, Aptdo.22085, 46071 Valencia, Spain}

\author{E. Oset}
\email[]{oset@ific.uv.es}
\affiliation{Departamento de F\'{\i}sica Te\'orica and IFIC, Centro Mixto Universidad de Valencia-CSIC Institutos de Investigaci\'on de Paterna, Aptdo.22085, 46071 Valencia, Spain}

\begin{abstract}
We address the issue of the compositeness of hadronic states and demonstrate that starting with a genuine state of nonmolecular nature, but which couples to some meson-meson component to be observable in that channel, if that state is blamed for a bound state appearing below the meson-meson threshold it gets dressed with a meson cloud and it becomes pure molecular in the limit case of zero binding. We discuss the issue of the scales, and see that if the genuine state has a mass very close to threshold, the theorem holds, but the molecular probability goes to unity in a very narrow range of energies close to threshold. The conclusion is that the value of the binding does not determine the compositeness of a state. However, in such extreme cases we see that the scattering length gets progressively smaller and the effective range grows indefinitely. In other words, the binding energy does not determine the compositeness of a state, but the additional information of the scattering length and effective range can provide an answer. We also show that the consideration of a direct attractive interaction between the mesons in addition to having a genuine component, increases the compositeness of the state. Explicit calculations are done for the $T_{cc}(3875)$ state, but are easily generalized to any hadronic system.

\end{abstract}

\maketitle

\section{Introduction}

The dilemma between molecular states and genuine quark states is the subject of a continuous debate in hadron physics. Concretely, concerning the  $T_{cc}(3875)$ state there are works that support the $T_{cc}$ as a molecular state  of $D D^{*}$ nature \cite{23,24,25,26,18,19,20,21,22,23bis,24bis,25bis,26bis,27,28,29,30,31,32,entem}, as well as others
 that advocate a  compact tetraquark nature \cite{27bis,28bis,29bis,30bis,32bis,33bis,taoguo,wzgang}, while other works suggest a mixture of both components \cite{yanparon,rosina}.

In the present work we start with a genuine state which allows to be observed in some meson-meson components and prove that in the
limit of small binding the state becomes purely molecular. The  issue of quark cores being dressed by molecular components is well
known and already discussed  in the past concerning the nature of the $``\sigma"$ meson ($f_0(500)$ nowadays)\cite{beveren1,beveren2,tronqvist}.
The dressing of a possible compact $T_{cc}$ state with $D D^{*}$ components
is also addressed in \cite{juanalba}.

We investigate in detail the scale, of what $``small~ binding''$ means to claim a full molecular state, and
show that the binding itself does not allow one to conclude that a state is molecular. On the other hand we also show that
if a pure genuine state is associated to a weakly bound state, it results into a very small scattering length and very
large effective range for the meson-meson component, which indicates that the measurement of these magnitudes is extremely useful to find out the nature of the hadronic states. In this respect it is  useful to call the attention to other works done in this direction.
In \cite{hyodotom} the compositeness (molecular probability) of hadronic state is discussed in terms of the binding, but
the consideration of the range of the interaction has as a consequence a larger molecular components for the $T_{cc}$ when
the range is changed from the long range of pion exchange to a shorter range of vector meson exchange. Probabilities of the
molecular component for only the $D^0 D^{*+}$ component are also evaluated in \cite{mishacom}. A more complete work considering
the scattering lengths and effective ranges, as well as the $D^0 D^0 \pi^+$ mass distribution of the experiment \cite{lhcbmisha,mishacom}, is done in \cite{ourwork} and
concludes that the sum of probabilities the $D^0 D^{*+}$ and $D^+ D^{*0}$ components,  is compatible with unity, stressing the  molecular nature of the state.
The value of the effective range and scattering length to determine the compositeness of  a state has also been emphasized
from the very beginning in the pioneer work of Weinberg \cite{weinberg} under strict conditions of zero range interaction and
very small binding,  but the first condition  was released in a recent work \cite{juancompo} and both conditions were released in
 the work of \cite{daisongcompo}, leading in both cases to strategies based on the knowledge of the binding, effective  range and scattering length that improve considerably over the original formulas of \cite{weinberg} (see also \cite{22,fkguo,baru,kinu}).

 The formalism presented here and the conclusions are general, but we  particularize to the study of the  $T_{cc}(3875)$  and
 show that the large effective range and scattering length that one obtains assuming a genuine state to be responsible for the
 $T_{cc}$  binding are very far off from those already determined from  the experimental study of this state.

\section{Formalism}

\begin{figure}[h!]
\centering
\includegraphics[scale=.9]{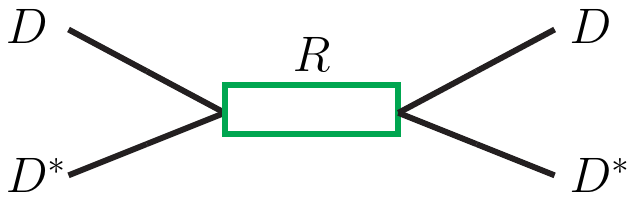}
\caption{ $D D^*$  amplitude based on the genuine resonance $R$.}
\label{fig:1}
\end{figure}

Let us  assume that we have a hadronic state of bare mass $m_R$, not generated by the interaction of meson-meson components, for instance a compact quark state.
We assume that even if small, the state couples to  one meson-meson component,  where the effects of this state can be observed. We think from the beginning  on the
$T_{cc}(3875)$ and the $D D^*$ component. To simplify the study we consider an $I=0$ state and just one channel, although the consequences are general and would apply
 to the lowest threshold of the   $D^0 D^{*+}$ component. This said, we can write for the $D D^*$  amplitude the diagram of Fig. \ref{fig:1}  and
the $D D^*$  amplitude of Eq.~\eqref{eq:t}.

\begin{eqnarray}\label{eq:t}
\tilde{t}_{D D^*,D D^*} (s)=\frac{\tilde{g}^2}{s-s_R}
\end{eqnarray}

\begin{figure}[h!]
\centering
\includegraphics[scale=.9]{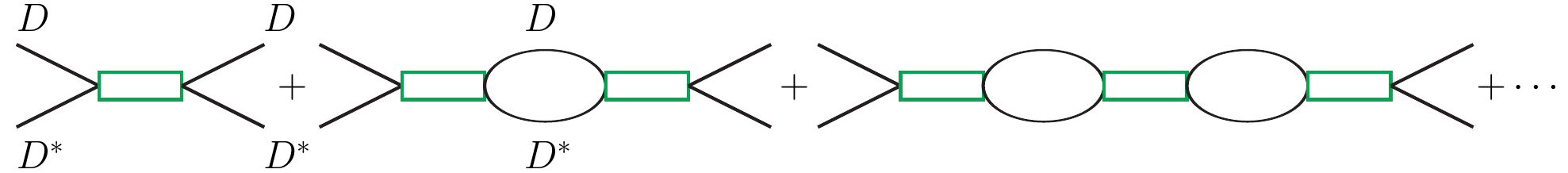}
\caption{Iterated diagram of Fig. \ref{fig:1} implementing unitarity of the $D D^*$  amplitude. }
\label{fig:2}
\end{figure}

This amplitude is not unitarity. It is rendered unitary immediately by iterating the diagram  of  Fig. \ref{fig:1} as shown in Fig. \ref{fig:2}. What we are doing
with the diagram of Fig. \ref{fig:2} is to insert the $D D^*$ selfenergy in the propagator of Eq.~\eqref{eq:t}. We have then
 \begin{eqnarray}\label{eq:t1}
t_{D D^*,D D^{*}}(s)=\frac{\tilde{g}^2}{s-s_R-\tilde{g}^2 G_{D D^*}(s) }
\end{eqnarray}
 where $G_{D D^*}(s)$ is the  $D D^*$ selfenergy which we choose to regularize with a sharp cutoff.
 \begin{eqnarray}\label{eq:cut}
G_{D D^*}(s)= \int_{|{\bm q}|<q_{\rm max}} \frac{d^3q}{(2\pi)^3} \, \frac{\omega_1 + \omega_2}{2 \,\omega_1 \,  \omega_2} \,\frac{1}{s-(\omega_1 + \omega_2)^2+i\epsilon}
\end{eqnarray}
where $\omega_i = \sqrt{{\bm{q}}^2 +m_i^2}$. The unitarity of the $t_{D D^*,D D^{*}}$ amplitude is shown immediately by means of
\begin{eqnarray} \label{eq:t2}
{\rm Im} \, t^{-1}={\rm Im} \left(\frac{s-s_R}{\tilde{g}^2}- G_{D D^*}(s)\right) =- {\rm Im} \, G_{D D^*}(s) = \frac{k}{8\pi \sqrt{s}}
\end{eqnarray}
with $k$ the meson-meson on shell momentum, $k=\lambda^{1/2}(s,m_D^2,m_{D^*}^2)/(2\sqrt{s})$.
Having $\tilde{g}^2$ positive and  ${\rm Re} \,G_{D D^*}(s)$ negative,
one can see  from Eq.~\eqref{eq:t1} that the
$D D^*$ selfenergy  is negative and moves the pole  $s_R$ of the bare resonance to lower energies. Let us  assume that
$\tilde{g}^2$ is such that the bare state $R$, conveniently dressed with the $D D^*$ selfenergy, is responsible for the
appearance of a pole at $s_0$, below the $D D^*$ threshold. Since the  $D D^*$ selfenergy is negative, we take then $s_R$ above the $D D^*$ threshold. Studies of the tetraquark structure for the  $T_{cc}$ state provide in most cases
masses above that threshold, like the one of Ref. \cite{quigg} which is $102\mev$ above the  $D^0 D^{*+}$ threshold,
and which we take as reference.

The condition that a pole appears at $s_0$ is easily obtained from Eq.~\eqref{eq:t1} as
\begin{eqnarray}\label{eq:pole}
s_0-s_R-\tilde{g}^2 G_{D D^{*}} (s_0)=0\,,
\end{eqnarray}
which provides the value of $\tilde{g}^2$ needed to accomplish it.

The next step is to calculate the molecular probability. According to \cite{hyodoijmp, danijuan} the molecular
  probability is obtained from
\begin{eqnarray}\label{eq:p}
P = -g^2\, \frac{\partial G}{\partial s}\big|_{s=s_0}
\end{eqnarray}
where $s_0$ is the square of the mass of the physical state, which we assume to be below the threshold, as in the case of the  $T_{cc}(3875)$.

In Eq.~\eqref{eq:p} $g$ is the coupling of the state to the  $D D^*$  component and $g^2$ the residue of the
$t_{D D^*,D D^{*}}$ matrix of  Eq.~\eqref{eq:t1}  at the pole. Thus
\begin{eqnarray}\label{eq:g2}
g^2 =  \lim_{s \to s_0} (s-s_0) \frac{\widetilde{g}^2}{s-s_R-\widetilde{g}^2 G_{D D^{*}}(s)}
= \frac{\widetilde{g}^2}{1-\widetilde{g}^2 \frac{\partial G}{\partial s} }\big|_{s=s_0}
\end{eqnarray}
where in the last step we have used L'H\^{o}pital rule. Then the molecular   probability is
\begin{eqnarray}\label{eq:p2}
P=- \frac{\widetilde{g}^2 \frac{\partial G}{\partial s}}{1-\widetilde{g}^2 \frac{\partial G}{\partial s} }\big|_{s=s_0}
\end{eqnarray}

We can see several limits:
\begin{align}
\begin{cases}
\tilde{g}^2 \to 0\,, \quad P \to 0 \,,  \quad  {\rm the~ genuine ~ state~ survives } \\
\tilde{g}^2 \to \infty\,,  \quad  P \to 1  \,,  \quad {\rm the ~state~ becomes~ pure ~ molecular }  \\
s_0 \to s_{\rm th}\,,  \quad  P \to 1 \,,  \quad  {\rm the ~state~ becomes~ pure ~ molecular }
 \end{cases} \nonumber
\end{align}

The third case is interesting, it is a consequence of unitarity and analyticity of the $t$ and $G$ functions. Indeed,
$\frac{\partial G}{\partial s} \to \infty/_{s_0 \to s_{\rm th}}$, and then the 1 in the
denominator of Eq.~\eqref{eq:g2} can be neglected and $P \to 1$. We can then state clearly that when the binding
energy goes to zero the state becomes fully molecular, the genuine  component has been  fagocitated  by the
molecular component that assumes all the  probability of the state.
This conclusion has also been reached before in \cite{juanalba,sazdjian}.
One might finish here, but there is the important issue of the scales.
In other words, what does $s_0 \to s_{\rm th}$ means in a real case, $10\mev$, $1\mev$, $10^{-2}\mev$? The answer to this
question is provided in the following section.

\section{Results for the compositeness as a function of $S_R$}

\begin{figure}[h!]
\centering
\includegraphics[scale=.75]{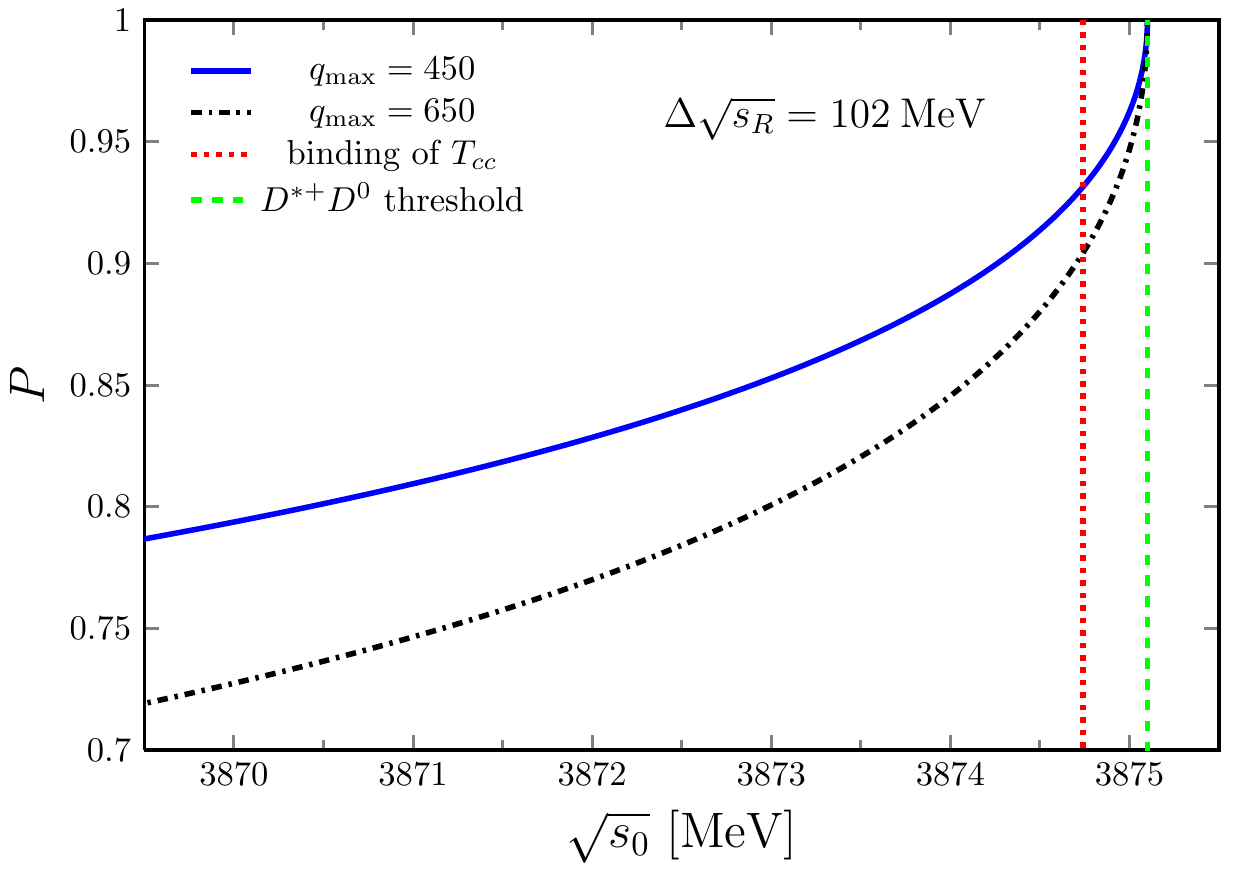}
\caption{Molecular  probability   based on the genuine resonance  with $\Delta\sqrt{s_R}=102\mev$.}
\label{fig:pro102}
\end{figure}

\begin{figure}[h!]
\centering
\includegraphics[scale=.75]{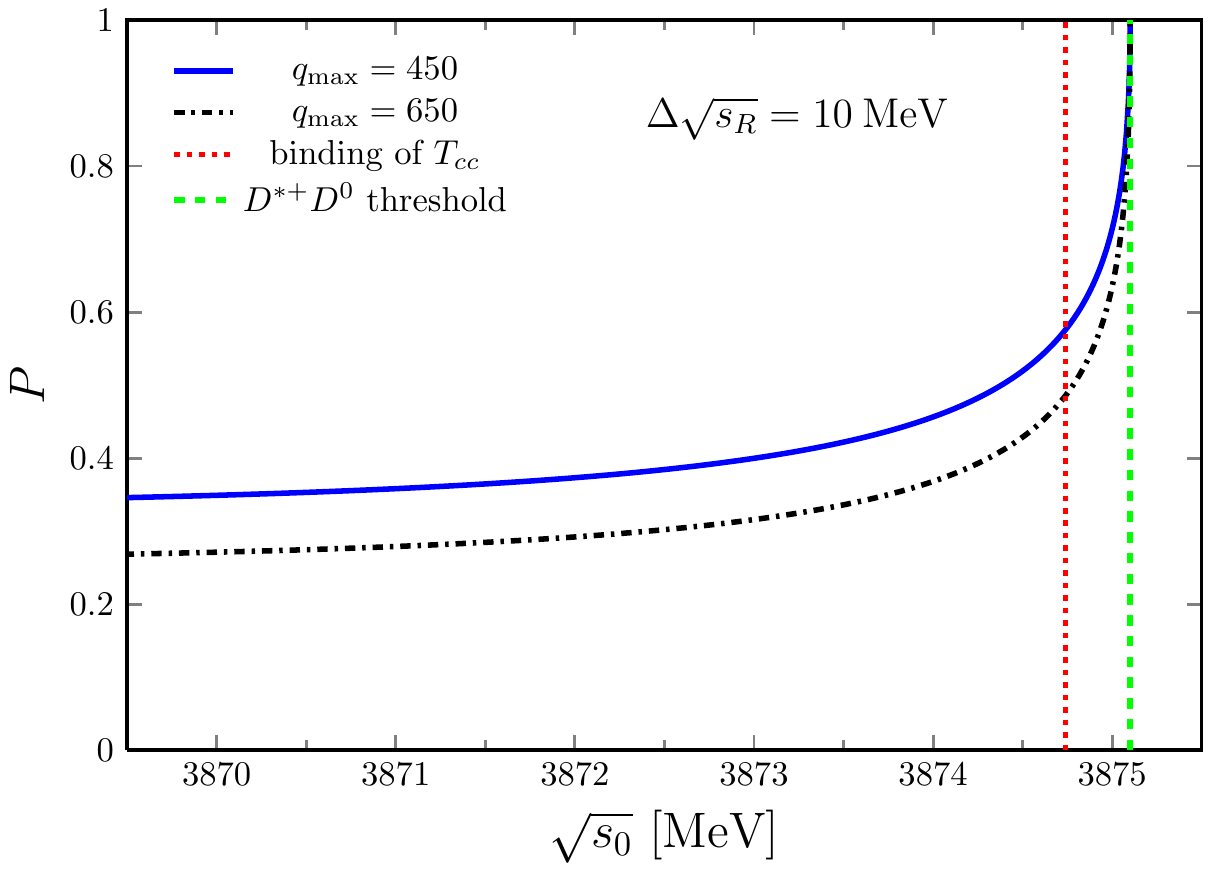}
\caption{Same as Fig.~\ref{fig:pro102} but  with $\Delta\sqrt{s_R}=10\mev$.}
\label{fig:pro10}
\end{figure}

\begin{figure}[h!]
\centering
\includegraphics[scale=.75]{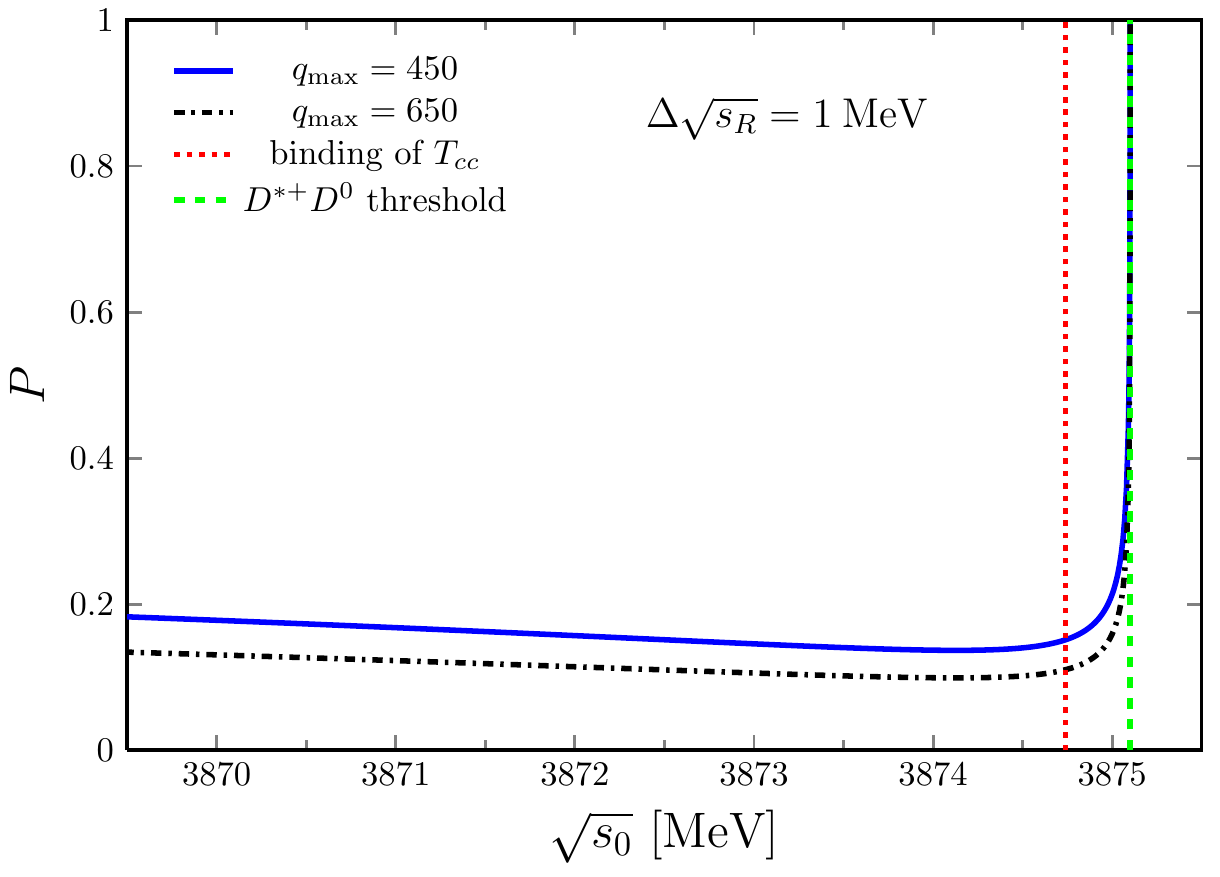}
\caption{Same as Fig.~\ref{fig:pro102} but  with $\Delta\sqrt{s_R}=1\mev$.}
\label{fig:pro1}
\end{figure}

\begin{figure}[h!]
\centering
\includegraphics[scale=.75]{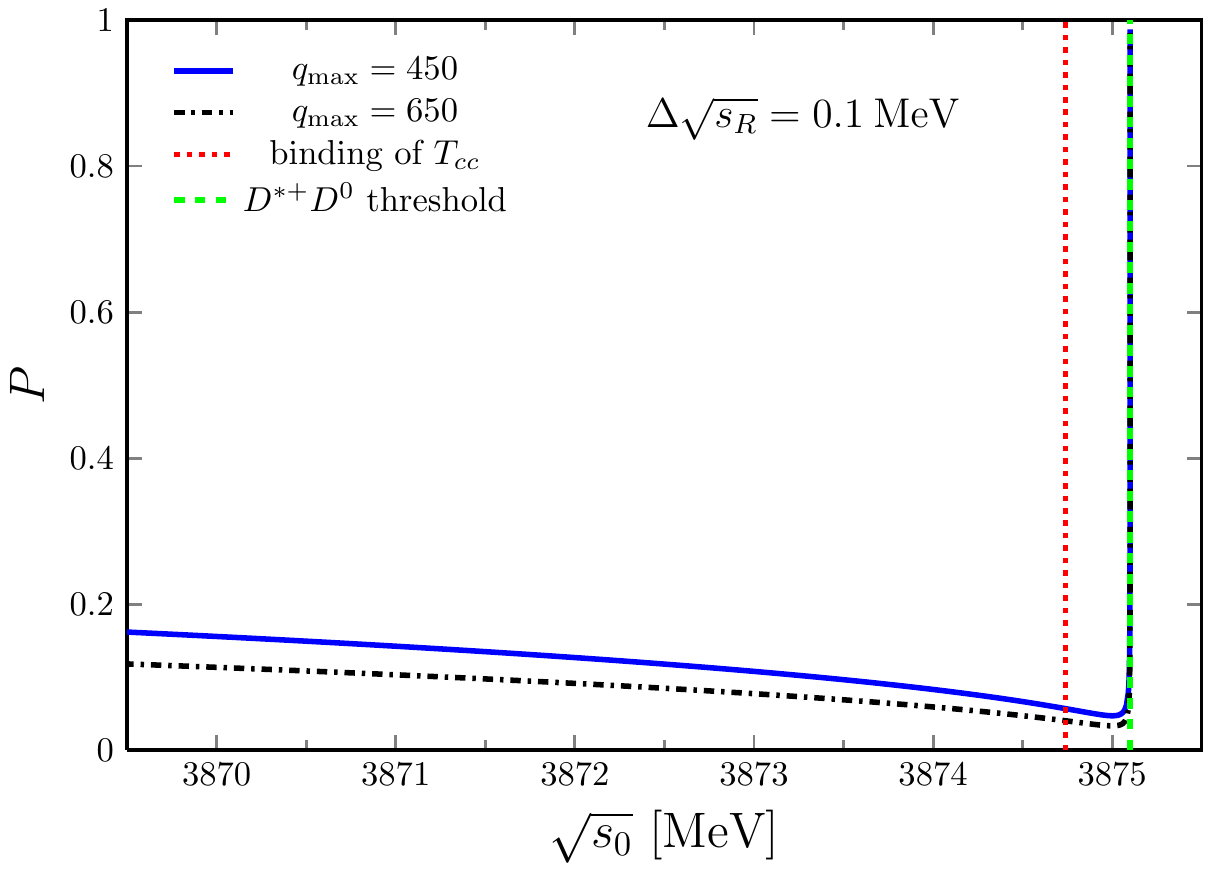}
\caption{Same as Fig.~\ref{fig:pro102} but  with $\Delta\sqrt{s_R}=0.1\mev$.}
\label{fig:prop1}
\end{figure}

In Figs.~\ref{fig:pro102}, \ref{fig:pro10}, \ref{fig:pro1},  \ref{fig:prop1} we show the results for the  molecular probability $P$
of Eq.~\eqref{eq:p2} for  different values of $s_R$, $s_R=\sqrt{s_{\rm th}}+\Delta\sqrt{s_R}$ with $\Delta\sqrt{s_R}=102,10,1,0.1\mev$,
as a function of $s_0$, the assumed value of the square of the energy of the bound state. In   Fig.~\ref{fig:pro102} we observe that for
 $\Delta\sqrt{s_R}=102\mev$,  $P$ goes indeed to $1$ when $s_0 \to s_{\rm th}$, as it should, but for $s_0^{\rm exp}$ $(\sqrt{s_0}=\sqrt{s_{\rm th}}-0.360\mev)$
 $P$ already has value around $0.9$, depending a bit on the assumed value of $q_{\rm max}$, indicating that the
 original genuine state has evolved to become practically a molecular state.

 The case of $\Delta\sqrt{s_R}=10\mev$ is shown in  Fig.~\ref{fig:pro10}. The trend is the same. $P \to 1 $ as $s_0 \to s_{\rm th}$, but
 for $s_0^{\rm exp}$ the value of $P$ is now smaller than before, of the order of $0.5$.

We repeat the calculations for $\Delta\sqrt{s_R}=1\mev$ in Fig.~\ref{fig:pro1} and we see now the same trend of $P$ when $s_0 \to s_{\rm th}$.
However, the ``scale" that we mentioned before shows up clearly since the change of  $P \to 1$ appears for values of $\sqrt{s_0} -\sqrt{s_{\rm th}}$ of the order of
$10^{-1}\mev$. For $s_0^{\rm exp}$ the value of  $P$  is smaller than $0.15$, indicating that the state remains mostly nonmolecular.

The results with the extreme  case of  $\Delta\sqrt{s_R}=0.1\mev$ further illustrate the point since now $P \to 1$ in an extremely narrow region of
 $s_0 \to s_{\rm th}$ and at  $s_0$  the value of $P$ is smaller than $0.05$. The state is basically nonmolecular in nature.

The results shown above indicate that the  value of the binding energy by itself cannot give a proof of the nature of the state.
Even if a state is very close to threshold, a genuine state with energy very close to threshold can reproduce the binding with a negligible  probability of
 molecular component. It is important to state this fact  because intuitively, a bound state very close to a threshold of a pair of particles is often interpreted
 as been a molecular state of that pair.

 This said, let us see what other magnitudes  can really tell us about the nature of the state.

\section{Scattering length and effective range}

The relationship of the scattering matrix $t$ with the one used  in Quantum Mechanics is given by
\begin{eqnarray}\label{eq:t3}
t=-8 \pi \sqrt{s} \, f^{\rm QM} \simeq -8 \pi\sqrt{s}  \, \frac{1}{-\frac{1}{a} + \frac{1}{2}\, r_0 \,k^2-ik}
\end{eqnarray}
then
 \begin{eqnarray}
 t^{-1}=-\frac{1}{8 \pi \sqrt{s} } \left(-\frac{1}{a}+ \frac{1}{2}\, r_0 \, k^2-ik\right)
\end{eqnarray}

Note that ${\rm Im} \, t^{-1}$ given by $ - {\rm Im} \, G_{D D^*}(s)$ in Eq.~\eqref{eq:t2} provides indeed the imaginary part of the right hand side of Eq.~\eqref{eq:t3}, the token of unitarity in the amplitude that we are using.  From
Eq.~\eqref{eq:t3} it is easy to induce
 \begin{eqnarray}
 -\frac{1}{a}=\frac{s_{\rm th}-s_R}{\tilde{g}^2}- {\rm Re} \, G_{D D^*} (s_{\rm th})
\end{eqnarray}
 \begin{eqnarray}
 \frac{1}{2} r_0=\frac{\partial}{\partial k^2} \left\{(-8 \pi \sqrt{s})\left(\frac{s-s_R}{\tilde{g}^2} -
  {\rm Re} \, G_{D D^*} (s)\right)\right\} \big|_{s=s_{\rm th}} \nonumber
\end{eqnarray}
or
 \begin{eqnarray}
r_0= 2 \frac{\sqrt{s}}{\mu}\frac{\partial}{\partial s} \left\{(-8 \pi \sqrt{s})\left(\frac{s-s_R}{\tilde{g}^2} -
  {\rm Re} \, G_{D D^*} (s)\right)\right\} \big|_{s=s_{\rm th}}
\end{eqnarray}
with  $\mu$ the reduced mass of the $D, D^*$ mesons with $\mu=m_D m_{D^*}/(m_D +m_{D^*})$.

In Table \ref{tab:1} we show the results of $a$ and $r_0$ as a function of  $\Delta\sqrt{s_R}$ when the state is bound at $s_0^{\rm exp}$. What we obtain is that as   $\Delta\sqrt{s_R}$  becomes smaller, decreasing the
molecular  probability, the scattering length becomes smaller and smaller and the effective range grows indefinitely. The values obtained for $\Delta\sqrt{s_R}=0.1\mev$, where the molecular  component is small, less than $0.05$,
are of the order of $0.61-0.87\fm$ for the scattering length, and of the order of  $-114 -(-168)\fm$.  Even for $\Delta\sqrt{s_R}=1\mev$ where the molecular  probability would be of the order of $15\%$, the scattering lengths
are in the range of $1.56-2.1\fm$ and the effective range from $-56.7 -(-38.2)\fm$. The lesson we draw from there is that
the values of $a$ and $r_0$  are very useful to determine the molecular  probability of the state. The numbers mentioned before are in sheer disagreement from those obtained experimentally in \cite{mishacom,lhcbmisha}, which are of the order of $a\sim 6-7\fm$, $r_0\sim -3.9\fm$ for the $D^0 D^{*+}$ channel. Let us stress once more that in the work of \cite{ourwork} the scattering length and effective range of the  $D^0 D^{*+}$, $D^+ D^{*0}$ channel, together with the
$D^0 D^0 \pi^+$ mass spectrum, were analyzed allowing both a molecular and a genuine component and it was concluded that
the state was $100\%$ molecular within the small uncertainties of the analysis. The present work offers a broad perspective on why that conclusion was obtained.

  \begin{table}[ht!]
\renewcommand\arraystretch{1.5}
\caption{\label{tab:1}  The obtained scattering length and effective range.  }
\begin{tabular*}{\columnwidth}{@{\extracolsep\fill}ccccc}
\hline\hline
 \multicolumn{1}{c}{\multirow{2}{*}{$\Delta \sqrt{s_R}$ [MeV]}} &  \multicolumn{2}{c}{$q_{\rm max}=450\mev$}  & \multicolumn{2}{c}{$q_{\rm max}=650\mev$}\\
\cline{2- 3} \cline{4-5}
  \multicolumn{1}{c}{}& $a$ [fm] & $r_{0}$ [fm]    &    $a$ [fm] & $r_{0}$ [fm]  \\
\hline
0.1  & 0.87 &   -114.07 &  0.61 & -168.39 \\
0.3 & 1.19&  -79.33  & 0.85 & -117.23\\
1  & 2.10 & -38.20     &  1.56 & -56.68 \\
2  & 3.04  & -21.77      & 2.36 & -32.49    \\
5  &   4.62 & -9.26    & 3.85 & -14.07 \\
10 & 5.74 & -4.51  & 5.07 & -7.08 \\
30 & 6.94 &  -1.16  & 6.54 & -2.14  \\
50 & 7.25 & -0.47  & 6.95 & -1.13\\
70  & 7.39 & -0.17    &  7.15 & -0.69\\
102 & 7.51 & 0.06  &  7.31 & -0.34\\
\hline\hline
\end{tabular*}
\end{table}

\section{Mixture of compact and molecular components}

So far we have just started from a pure nonmolecular state and we show that the  dressing with the meson-meson
cloud renders the state molecular in the limit of a small binding. The pure molecular states are obtained starting
with an energy independent potential $V$ between the particles of the meson pair, with the  scattering amplitude
becoming
\begin{eqnarray}\label{eq:t4}
T=\frac{V}{1-V G}
\end{eqnarray}

If we have a mixture of the genuine  state and the molecular one, this can be accounted for by taking a
potential
\begin{eqnarray}
V'=V+\frac{\tilde{g}^2}{s-s_R}
\end{eqnarray}

It is easy to generalize the probability $P$ to this case and we find
\begin{eqnarray}\label{eq:p3}
P=- \frac{  \left[\widetilde{g}^2+(s-s_R)V \right]  \frac{\partial G}{\partial s}}{1- \left[\widetilde{g}^2+(s-s_R)V \right] \frac{\partial G}{\partial s}-VG }\big|_{s=s_0}
\end{eqnarray}

The pole at $s_0$ appears now when
 \begin{eqnarray}
 s_0-s_R-\left[\widetilde{g}^2+(s_0-s_R) V \right] G(s_0)=0
\end{eqnarray}

We conduct now a new test. We take a potential $V$ short of binding, meaning that by itself would have
$1-V G(s)$ of the denominator of Eq.~\eqref{eq:t4} at the threshold $s=s_{\rm th}$. Hence
\begin{eqnarray}
 1-V G (s_{\rm th})=0
\end{eqnarray}
We compare this  potential  with the one we obtain from the local hidden gauge approach
\cite{h1,h2,h3,h4}
 \begin{eqnarray}
 V =\beta V_{\rm LHG} =\beta \,(-1) \,\frac{1}{2} \, g'^2 \left[3s-(M^2+m^2+M'^2+m'^2)-\frac{1}{s}(M^2-m^2)(M'^2-m'^2)\right] \frac{1}{M^2_{\rho}}
\end{eqnarray}
with $g'=\frac{M_V}{2\,f} ~(M_V=800\mev, f=93\mev)$ and $M,m$ the masses of $D^*$ and $D$, and
the same for $M',m'$. We obtain
\begin{align}
\begin{cases}
\beta=0.74    \quad  {\rm for~} q_{\rm max}=450\mev  \\[0.1cm]
\beta=0.52   \quad  {\rm for~} q_{\rm max}=650\mev
 \end{cases} \nonumber
\end{align}
Since $V$ is short of binding, we allow the nonmolecular component, the term $\tilde{g}^2/(s-s_R)$ to be responsible for the binding. Then we obtain the results of $P$ shown in Table \ref{tab:2}.

\begin{table}[ht!]
\renewcommand\arraystretch{1.5}
\caption{\label{tab:2} The molecular  probability $P$ of the state. }
\begin{tabular*}{\columnwidth}{@{\extracolsep\fill}ccccc}
\hline\hline
 \multicolumn{1}{c}{\multirow{2}{*}{$\Delta \sqrt{s_R}$ [MeV]}} &  \multicolumn{2}{c}{$q_{\rm max}=450\mev$}  & \multicolumn{2}{c}{$q_{\rm max}=650\mev$}\\
\cline{2- 3} \cline{4-5}
  \multicolumn{1}{c}{}& $\beta=0$ & $\beta=0.74$   &    $\beta=0$ & $\beta=0.52$  \\
\hline
10 & 0.58 & 0.94	&  0.49& 0.94 \\
20 &  0.73 & 0.97	  &  0.65 &0.97    \\
50 &  0.87   & 0.99  & 0.82 & 0.99  \\
\hline\hline
\end{tabular*}
\end{table}

For different values of $\Delta \sqrt{s_R}$ what we find is that if in addition to the genuine  state we add some
potential between the $D,D^*$ strong enough, but not enough to bind by itself, the effect of it is that it increases the molecular  probability bringing it close to unity.
We also observe the feature that the bigger the value of $s_R$, the smaller is the relative increase in the compositeness (see  also
similar results in related studies in connection  with lattice QCD data \cite{raq}).
What one concludes from here is that if one has a state close to threshold of a pair of particles and there is some attractive interacting potential between  these particles, the
chance that the state is a molecular state increases appreciably. Certainly, if the potential is enough to bind by itself one does not need a nonmolecular  component, but what we see is that even if it exists it does not change
the fate of the state turning molecular. Yet, the complement of  the scattering length and effective range,
as well as mass distribution close to threshold, help finally to make a precise determination of the molecular  probability of the state.

\section{Conclusions}
   In this work we have addressed the issue of the dressing of an elementary, or genuine state, by meson components and how this genuine state can eventually turn into a pure mesonic molecular state due to this meson cloud. For this purpose we start from a state which is purely genuine, let us say for instance a compact quark state, which has a certain coupling to a meson-meson component, such that its effects can be observed in this meson-meson channel. Then we demand that this state becomes a bound state below the meson-meson threshold and then determine the probability that the state has become molecular.  We demonstrate that when the binding energy of the state goes to the meson-meson threshold, the state becomes $100\%$ molecular.  Yet, the important issue is the scale of energies where this happens.  We discuss the issue in detail. For this purpose we show the molecular probability as a function of the binding energy for different values of the genuine state mass, $M_R$. We observe that if $M_R$ is far away from the meson-meson threshold, then the bound state goes fast to being molecular as we approach the threshold. However, as $M_R$ gets closer to the meson-meson threshold the theorem holds equally but the probability goes only to $100\%$ at energies extremely close to threshold, such that even for states bound by $0.360$ MeV, like the $T_{cc}(3875)$, the molecular probability can be very small.  The conclusion is that the proximity of a state to a threshold is not a guaranty that the state is of molecular nature. However, there is a consequence of having the genuine state responsible for the state found, because the  scattering length becomes gradually smaller and the effective range grows indefinitely and reaches unphysical values for a case like the $T_{cc}(3875)$ mentioned above. Indeed, we find that if one demands that the $T_{cc}(3875)$ is a genuine, nonmolecular state, the scattering length and effective range obtained are in sheer disagreement with data.   The conclusion is then that the binding, together with measurements of the scattering length and effective range can provide an answer to the compositeness of a state, but not the binding alone.

   We also show that if we have a mixture of a genuine state and an additional direct attractive interaction between the mesons, the state becomes more molecular for the same mass $M_R$ of the genuine state. Certainly, with enough attraction, one can generate the state without the need of an extra genuine component.
   The present work brings light to the continuous debate over the nature of hadronic states and provides a perspective on issues discussed before in the Literature, on the relevance  of the scattering length and effective range, or mass distributions close to threshold, to determine the compositeness of hadronic states. Although we have particularized the calculations for the case of the  $T_{cc}(3875)$, the results and conclusions are general and the method employed in the analysis can be easily extrapolated to any other hadronic cases.



\section*{Acknowledgments}
This work is partly  supported by the National Natural Science Foundation of China
under Grants Nos. 12175066, 11975009, 12247108 and the China Postdoctoral Science Foundation under Grant No. 2022M720359.
This work is also partly supported by the Spanish Ministerio de
Economia y Competitividad (MINECO) and European FEDER funds under Contracts No. FIS2017-84038-C2-1-P
B, PID2020-112777GB-I00, and by Generalitat Valenciana under contract PROMETEO/2020/023. This project has
received funding from the European Union Horizon 2020 research and innovation programme under the program
H2020-INFRAIA-2018-1, grant agreement No. 824093 of the STRONG-2020 project.
This research is also supported by the Munich Institute for Astro-, Particle and BioPhysics (MIAPbP)
which is funded by the Deutsche Forschungsgemeinschaft (DFG, German Research Foundation)
under Germany's Excellence Strategy-EXC-2094 -390783311.

\end{document}